\documentclass[conference]{IEEEtran}
\IEEEoverridecommandlockouts
\usepackage{cite}
\usepackage{amsmath,amssymb,amsfonts}
\usepackage{algorithmic}
\usepackage{graphicx}
\usepackage{textcomp}
\usepackage{multicol, multirow}
\usepackage{xcolor}
\usepackage{comment}
\usepackage{hyperref}
\def\BibTeX{{\rm B\kern-.05em{\sc i\kern-.025em b}\kern-.08em
    T\kern-.1667em\lower.7ex\hbox{E}\kern-.125emX}}
\begin{document}

\title{Agent-based Simulation for Drone Charging in an Internet of Things Environment System \\

\thanks{This study was financed in part by the Coordenação de Aperfeiçoamento de Pessoal de Nível Superior—Brasil (CAPES)—Finance Code 001.}
}

\author{\IEEEauthorblockN{1\textsuperscript{st} Leonardo Grando}
\IEEEauthorblockA{\textit{University of Campinas (UNICAMP)} \\
Limeira, Brazil \\
https://orcid.org/0000-0002-0448-209X}
\and
\IEEEauthorblockN{2\textsuperscript{nd} Jose Roberto Emiliano Leite}
\IEEEauthorblockA{\textit{University of Campinas (UNICAMP)} \\
Limeira, Brazil \\
https://orcid.org/0000-0001-8368-7386}
\and
\IEEEauthorblockN{3\textsuperscript{rd} Edson Luiz Ursini}
\IEEEauthorblockA{\textit{University of Campinas (UNICAMP)} \\
Limeira, Brazil \\
https://orcid.org/0000-0002-1597-4057}}

\maketitle

\begin{abstract}
This paper presents an agent-based simulation model for coordinating battery recharging in drone swarms, focusing on applications in Internet of Things (IoT) and Industry 4.0 environments. The proposed model includes a detailed description of the simulation methodology, system architecture, and implementation. One practical use case is explored: Smart Farming, highlighting how autonomous coordination strategies can optimize battery usage and mission efficiency in large-scale drone deployments. This work uses a machine learning technique to analyze the agent-based simulation sensitivity analysis output results. 
\end{abstract}



\section{Introduction}

Drones have become important tools within the Internet of Things, and can be used in agribusiness, disaster response, logistics, and other usages. Their flight capacity due to energy limitation is one of the most influential barriers to drone applications in this context \cite{Rahman2021}.

This model explores how this limitation can be simulated and studied. In the IoT layer model, drones are in the auxiliary equipment layer for data collection (photos, images, temperature, humidity, etc.). 

\begin{figure}[ht]
	\centering
	\includegraphics[width=0.3\textwidth]{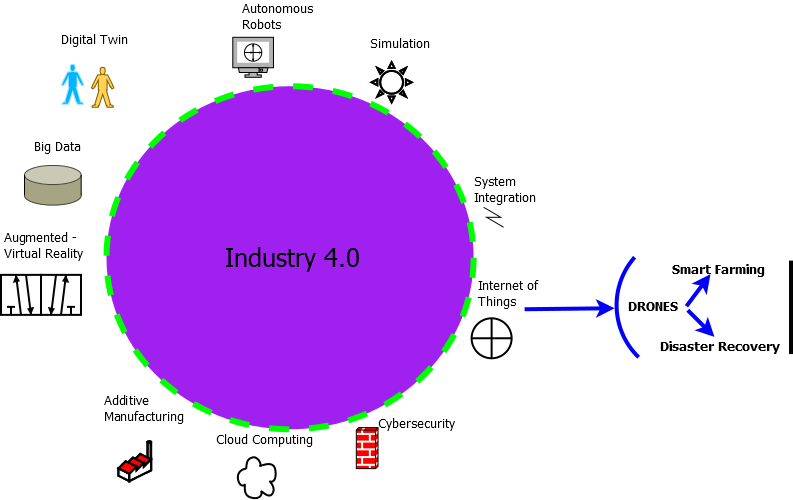} \\
    \caption{Industry 4.0 model based on \cite{Leite2023}.}
	\label{fig:1}
\end{figure}

Figure \ref{fig:1} presents an Industry 4.0 model, with several actors including autonomous robots (drones), agent-based simulation, the Internet of Things, systems integration, and others \cite{Leite2023}.

\begin{figure}[ht]
	\centering
	\includegraphics[width=0.3\textwidth]{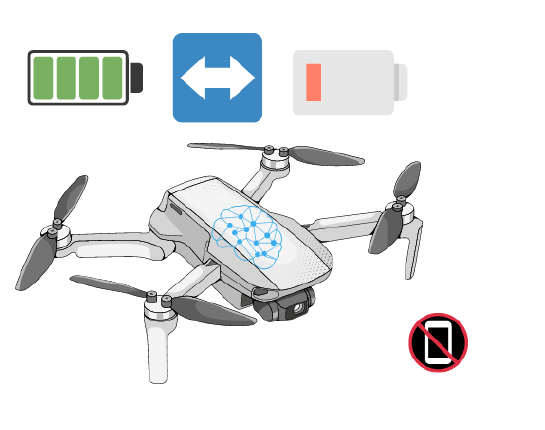} \\
    \caption{This work model idea.}
	\label{fig:2}
\end{figure}

This work aims to study using agent-based simulation (ABS), a swarm of drones' battery charging coordination using a decentralized approach to their recharging decision, as it is a critical part of their usage time, as shown in Figure \ref{fig:2}.

This article's structure is: Section \ref{literature} presents the characteristics and trends of drone technology and agent-based simulation. Section \ref{methodology} addresses the IoT Applications that are the main focus of this work: smart farming, disaster recovery, and fighting dengue, showing their current needs, developments, and trends. Section \ref{results} introduces the agent-based modeling and simulation of recharging drone batteries and details of the Work Development: location, variables, code, and environment analysis. Section \ref{conclusions} describes the results and discussions about the values obtained. Finally, Section 6 presents conclusions and considerations, research limitations, and suggestions for future work.

\section{Literature} \label{literature}

\subsection{IoT Architectures, Models, Protocols, Technologies, and Applications}       

The IoT Architecture presents an overview of the IoT: Architectures, Models, Protocols, Technologies, and Applications \cite{ITU-T2016,Leite2019c}. 

The IoT Architecture chooses open, simple, cheap, and reliable technologies such as wireless and wired as Ethernet, WiFi, Ad-Hoc, ZIGBEE, Bluetooth, Radio-Frequency Identification (RFID), Wireless Signal Solutions (WSS), 5G/6G mobile network, and sensors. 

\begin{figure}[ht]
	\centering
	\includegraphics[width=0.4\textwidth]{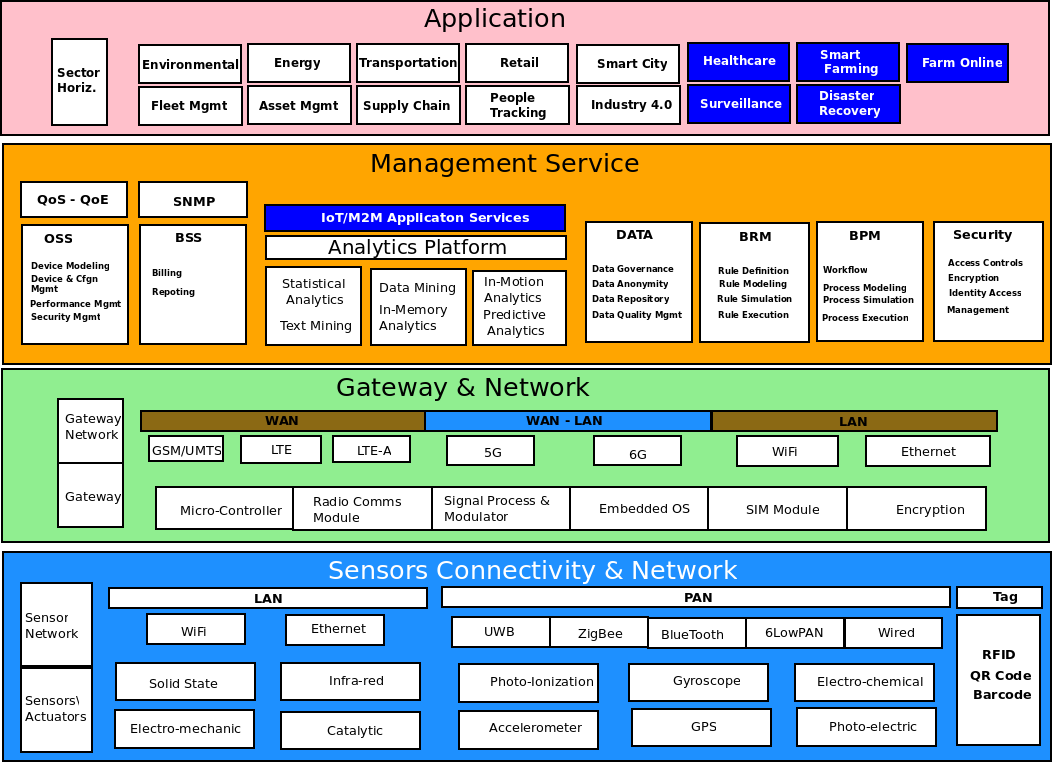} \\
    \caption{IoT Architecture based in \cite{Patel2016,Leite2019c}.}
	\label{fig:3}
\end{figure}

Figure \ref{fig:3} presents an IoT Architecture, selected from other established Architectures such as TCP/IP and SOA (Service Oriented Architecture). The IoT Architecture separates Communication into the following Layers: Sensor and Network Connectivity; Gateways and Networks; Network and Information Security Management Services, and Applications. Alongside LTE (4G), there is currently the 5G Mobile Network and, in the future, 6G \cite{Leite2019c}. 

The IoT made it possible to create applications for direct use in society and the World Economy, such as Environmental, Energy (Smart Grid), Transport, Health, Retail/Wholesale, Production Chain, Monitoring of People's Health (Healthcare ) and Security (Smart Home, Building, City), among others \cite{Patel2016}. 

\subsection{Selected applications in the IoT}

Regarding the drones recharging coordination process, a Systematic Literature Review in agriculture and disaster context \cite{Grando2025} found three research gaps about knowledge, methodological, and practical areas. Because of useful applications in these areas motivate us to perform this simulation work.

Smart Farming, or Precision Agriculture (PA), automates the farm through its interconnection with the global world. It enables the interconnection of tractors, harvesters, drones, farm headquarters, cooperatives, agricultural entities, and spray planes. Employees and animals can be identified using RFID tags \cite{Fathallah2017, Patel2016, Hartanto2019, RadoglouGrammatikis2020, Rahman2021}. 

\begin{figure}[ht]
	\centering
	\includegraphics[width=0.3\textwidth]{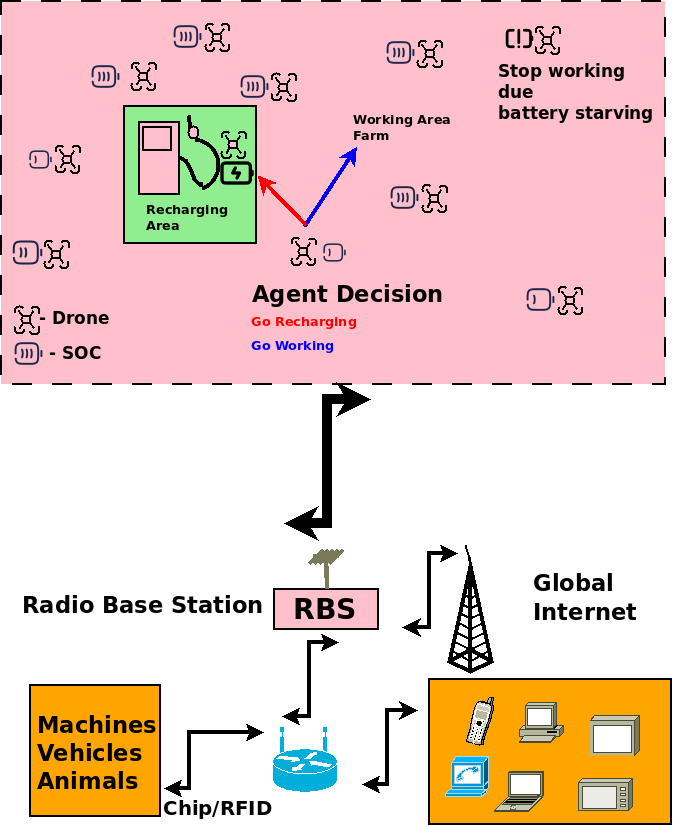} \\
    \caption{Our proposed PA framework.}
	\label{fig:5}
\end{figure}

Figure \ref{fig:5} presents how a Radio Base Station allows the interconnection of these different parts, which makes it possible to build a connected farm. This application is of great importance for Brazil as it is a heavily agricultural and livestock country \cite{FAO2022}.

\section{Methodology} \label{methodology}

\subsection{Model Simulation description}

In this model, the drones (agents), considered as autonomous agents. Agents (drones) are heterogeneous; they make decisions autonomously, but they do not interact with each other \cite{Macal2016}.

At first, we focused on two problems to be minimized by this strategy: 
\begin{enumerate}
    \item the reduction in communication between drones (which can occur in other critical situations) with the consequent reduction in battery consumption;
    \item Mitigate the need for remote control to recharge drones during continuous work on farms or disasters in the recharging process.
\end{enumerate}

This simulation model was deployed in the NetLogo \cite{Wilensky1999}, an agent-based simulation software. Our proposed model considers the agents' recharging coordination process without collusion. 

According to \cite{macal2005}, Agent-Based Simulation is built from a perspective from scratch, with an appropriate point of view, considering large-scale projects as complex systems, but having 1) agents that are part of the environment, directed towards the established objective, autonomous and that can learn and adapt, 2) local interaction through rules between agents, and 3) global configurations on the environment.

Our ABMS problem considers conditions 1) and 3), but condition 2) is replaced by the typical behavior of the El Farol Bar model, in which no communication between agents, but policies apply to the entire swarm of drones to carry out the mission.

The simulation objective is to develop a process for coordinating the decision to recharge a swarm of drones, where they will define whether or not to recharge their batteries. To make this decision, each agent has internal decision polices, which we call decision policies. In our proposal, the Charger Threshold (CT) policy uses the previous value of the number of drones that visited the charging station to make its decision. This policy is based on the El Farol Bar Problem \cite{Arthur1994}. 

This model considers a neighborhood with a certain number of agents $(QTY)$. They want to decide whether to go to this bar. This decision is based on the previous occupancy value history. The occupation limit uses a happiness criterion called the overcrowding threshold $B$. When the bar occupancy is lower than $B$, bar attendees will enjoy a good night. This agent uses a set of $k$ autoregressive predictors that use a window of previous values of the bar's occupancy in $m$ previous weeks to estimate the next agent's predicted attendance value. If the agent's predicted value is less than $B$, these agents will attend the bar in the next simulation round. 
 
Our model uses the El Farol NetLogo model \cite{Rand2007} library as a base for the simulation implementation. We use the same autoregressive model whose estimators' internal weights were defined at the beginning of the simulation. We also use this approach to calculate predicted values in our simulation model.

In our ABM, we consider the drones as agents, the charging station as the El Farol Bar, and the predictors as a source of calculations so that the agents' internal policies can make their decisions. The El Farol Bar Problem is already used in resource congestion problems \cite{Bell1999, Sharif2011}.

\subsection{Variables and Decision Process}

In this model, a $QTY$ number of drones decides which action to take when choosing between visiting the work environment or the charging station. The first location is where it performs its mission (spraying, obtaining images, serving as a source of connectivity in case of disasters), and the second place is its charging station. This charging station has a limited capacity, with this capacity being a fraction of the drone $QTY$ values.
The simulation model is related to the problem's demand, the system's recharge capacity (supply), and the agents' decision process.
Concerning the demand for the swarm of drones, they are related to the Quantity of Agents (Quantity - $QTY$), the average energy expenditure of each agent (Battery Consumption - $BC$), and the standard deviation of BC ($SD$).
Regarding the recharging offer for drones, the capacity of the recharging station $(B)$ and the amount of energy supplied with each recharge carried out (Battery Gain - $BG$) are the considered variables.

CT Policy considers both the amount of energy present in the drone battery and the decision process based on the El Farol Bar, the history of agents who went to recharge at the charging station to make a decision. Regarding the history of agents that have tried to recharge their batteries in the last m weeks, this information was sent by the Radio Base Station ($RBS$), where each agent would use these previous size values to the $k$ internal predictors and calculate the predicted value. If the drone's internal predicted value is less than $B$, the agent decides to recharge the batteries or continue its work.
CT policy also considers the battery values of these drones either at the lower value ($LW$) so that the agent who has a charge lower than this value decides to recharge, or the higher value ($UP$), the battery level value is higher than the UP, the drone will not try to recharge.

\begin{figure}[ht]
	\centering
	\includegraphics[width=0.3\textwidth]{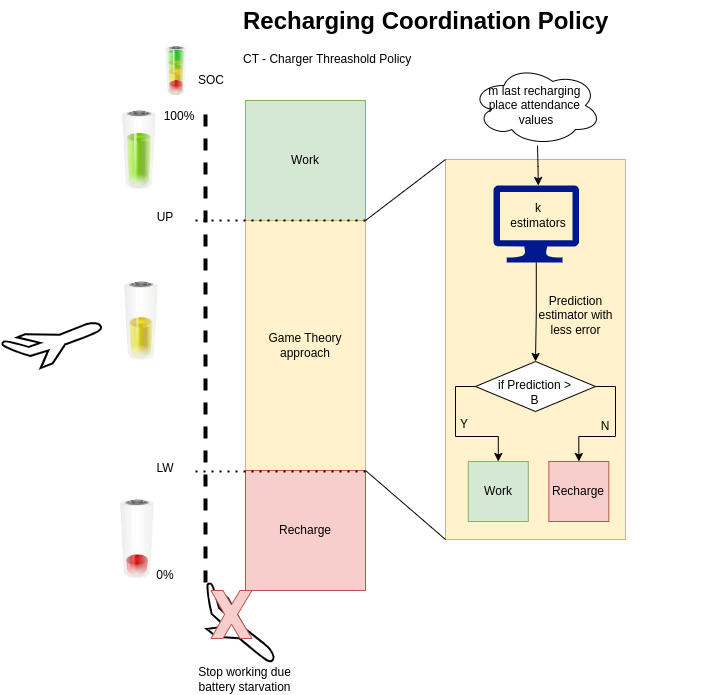} \\
    \caption{Recharge Policy Decision Process.}
	\label{fig:10}
\end{figure}

About the drone battery modeling, the battery State of Charge $(SOC)$ range goes from 0 to 100\% battery level. Two simulation parameters increase or decrease the $SOC$ during the simulation. The first is the Battery Gain $(BG)$ that increases the $SOC$, emulating the battery hot-swapping process in a given time (1 time step). 

In each run, the drones' $SOC$ value decreases at a Battery Consumption $(BC)$ rate. Because drone usage can be randomized by environment and work factors such as UAV payload, wind speed, and UAV vertical and horizontal movement, the model considers a normal standard deviation value $(SD)$ in the Battery Consumption value.
Regarding time dimension, we use Netlogo’s internal clock time step unit, called $tick$. Drones move randomly in a 2D space, according to their state defined during the simulation run, to recharging or working patches.

The experiment consists of a sensitivity analysis of the model depending on battery demand values. The experiment evaluates 9 parameters with 2 levels, with 10 repetitions, resulting in 5120 simulation rounds. Table \ref{tab:simulation_parameters} presents simulation parameter values. 

\begin{table}[ht]
\centering
\caption{Simulation parameters.}
\begin{tabular}{clcc}
\hline
 \textbf{Symbol} & \textbf{Description} & \textbf{-} & \textbf{+} \\
\hline 
  $m$   & Estimator Data windows.                      & 2  & 9 \\
  $k$   & Internal estimator numbers          & 2 &  9  \\
  $UP$    & Decision upper value                        & 70 & 90  \\
  $LW$    & Decision lower value               & 25  & 50 \\
  $BC$   & SOC Consumption in each run       & 10 & 15 \\
  $SD$    & BC standard deviation                        & 0 & 0.1  \\
  $QTY$   & The initial number of drones                 & 50 & 100  \\
  $BG$    & SOC increase in effective recharging         & 50 & 100 \\
  $B$      & Recharging place capacity & 30 & 60 \\
\hline
\end{tabular}
\label{tab:simulation_parameters}
\end{table}

The stopping criterion for each round is the presence of no agent with a remaining battery or 1500 simulation rounds. 

The experiments' results are concerning the reliability of each simulation set evaluated by the Average Simulation Remaining Drones ($ASRD$), related to the average number of drones that completed the simulations. It represents that a system will have great mission coverage. This indicator is evaluated by the ratio of the remaining simulation drones by $QTY$, as shown by Equation \ref{eq}.

\begin{equation}
ASRD = \frac{Simulation Remaining Drones}{QTY}\label{eq}
\end{equation}

An $ASRD = 1$ is a case when all drones finish the simulation, which means that it is a reliable system. 

This work have difference from \cite{grando_modeling_2024, Grando2020} works regarding the experiments performed evaluates more simulation parameters and simulation output evaluation techniques.
First, we will describe the $ASRD < 1$ statistical results, after evaluating which parameter is more important using the Random Forest Classifier \cite{scikit-learn} machine learning technique. 

\section{Results and discussion} \label{results}

\subsection{$ASRD$ Descritive Statistical Analisys}

There are 1109 cases from 5120 (21,7\%) of $ASRD<1$ cases. Table \ref{tab:descriptive_statistics} shows the summary statistics of these cases.

\begin{table}[ht]
\centering
\caption{Descriptive statistics of the $ASRD<1$ cases}
\begin{tabular}{l r}
\hline
\textbf{Metric} & \textbf{Value} \\
\hline
Mean            & 0.379 \\
Standard Deviation & 0.291 \\
Minimum         & 0.000 \\
1st Quartile (25\%) & 0.080 \\
Median (50\%)   & 0.370 \\
3rd Quartile (75\%) & 0.560 \\
Maximum         & 0.990 \\
\hline
\end{tabular}
\label{tab:descriptive_statistics}
\end{table}

\subsection{Variable importance to $ASRD < 1$}

\begin{figure}[ht]
	\centering
	\includegraphics[width=0.45\textwidth]{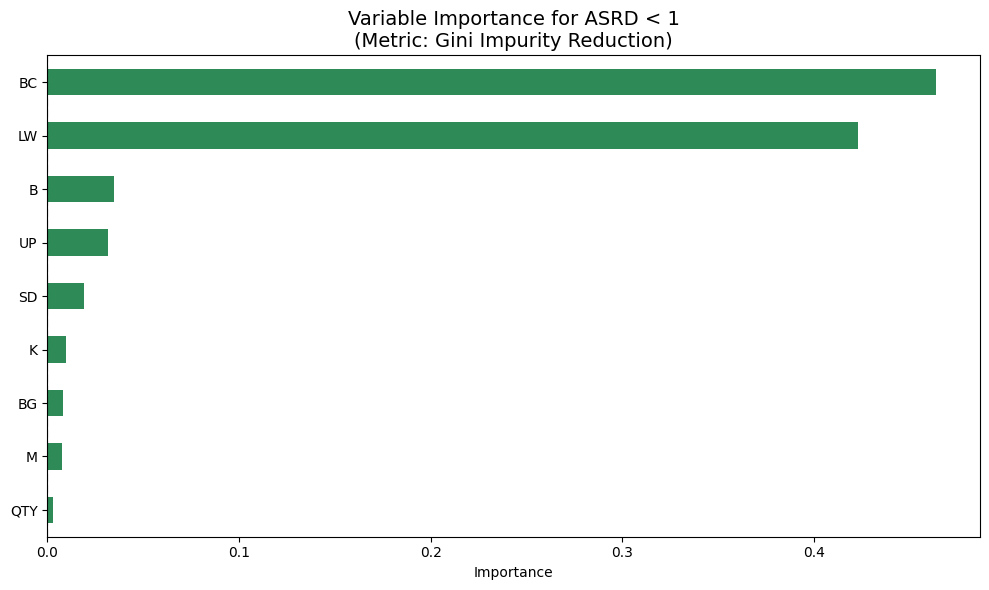} \\
    \caption{Variable importance to $ASRD < 1$}
	\label{fig:11}
\end{figure}

Figure \ref{fig:11} illustrates the variable importance scores obtained using a Random Forest classifier to predict the occurrence of critical $ASRD$ events. The metric used to compute variable importance was the Gini impurity reduction, which measures how effectively each variable contributes to class separation in the decision trees that compose the forest.

The results indicate that the variables $BC$ and $LW$ are the most influential factors, contributing the largest reductions in node impurity. Together, these two variables have a critical role in explaining the occurrence of $ASRD < 1$ cases. Other variables, such as $B$, $UP$, and $SD$, have a minor but non-negligible influence, while the remaining variables contribute minimally to the model.

The simulation model and the data output \href{https://github.com/lgrando1/SIoT-2025}{https://github.com/lgrando1/SIoT-2025}.

\section{Conclusions} \label{conclusions}

This work presents an Agent-Based Modeling Simulating a swarm of IoT devices in a high-energy-demand application. We create an agent-based simulation approach to evaluating how 512 simulation sets perform in several missions and which parameters are more influential in creating $(ASRD = 1)$ perpetual flight cases in a decentralized recharging drones swarm decision process using the $CT$ policy. 
We use a machine learning approach to evaluate how the nine parameters impact the simulation. Two parameters ($BC$ and $LW$) have the most influential impact and will be evaluated further in future works. Another's output indicators can be deployed in future works. 


\bibliographystyle{unsrt}  
\bibliography{template}

\end{document}